# Skyrmion Logic System for Large-Scale Reversible Computation

*Maverick Chauwin[1,2]‡, Xuan Hu[1]‡, Felipe Garcia-Sanchez[3,4], Neilesh Betrabet[1], Alexandru Paler[5], Christoforos Moutafis[6], & Joseph S. Friedman[1]\**

[1]Department of Electrical & Computer Engineering, The University of Texas at Dallas, Richardson, TX 75080, USA.

[2]Department of Physics, École Polytechnique, 91128 Palaiseau, France.

[3]Istituto Nazionale di Ricerca Metrologica, 10135 Torino, Italy.

[4]Departamento de Física Aplicada, Universidad de Salamanca, 37008 Salamanca, Spain.

[5]Linz Institute of Technology, 4040 Linz, Austria.

[6]School of Computer Science, The University of Manchester, Manchester M13 9PL, UK.

‡These authors contributed equally to this work.

\*Correspondence to: joseph.friedman@utdallas.edu.

ABSTRACT: **Computational reversibility is necessary for quantum computation and inspires the development of computing systems in which information carriers are conserved as they flow through a circuit. While conservative logic provides an exciting vision for reversible computing with no energy dissipation, the large dimensions of information carriers in previous realizations detract from the system efficiency, and nanoscale conservative logic remains elusive. We therefore propose a non-volatile reversible computing system in which the information carriers are magnetic skyrmions, topologically-stable magnetic whirls. These nanoscale quasiparticles interact with one another via the spin-Hall and skyrmion-Hall effects as they propagate through ferromagnetic nanowires structured to form cascaded conservative logic gates. These logic gates can be directly cascaded in large-scale systems that perform complex logic functions, with signal integrity provided by clocked synchronization structures. The feasibility of the proposed system is demonstrated through micromagnetic simulations of Boolean logic gates, a Fredkin gate, and a cascaded full adder. As skyrmions can be transported in a pipelined and non-volatile manner at room temperature without the motion of any physical particles, this skyrmion logic system has the potential to deliver scalable high-speed low-power reversible Boolean and quantum computing.**

## I. INTRODUCTION

There is a fundamental minimum quantity of energy dissipated by a logic gate in which information-carrying signals are continuously created and destroyed, as determined by Landauer [1]. Reversible computing aims to circumvent this limitation by conserving information – and therefore energy – as signals propagate through a logic circuit [2]. In this scheme, conservative logical operations are executed through dissipation-free elastic interactions among these information carriers that conserve momentum and energy [3]. While Fredkin and Toffoli's original thought experiment considered billiard balls as the information carriers, Prakash *et al.* recently demonstrated conservative logic experimentally with micron-sized droplets driven through planar computing structures by pressure [4] and magnetism [5]. However, the large dimensions of the information carriers in these demonstrations detract from the system efficiency and limit potential utility; a nanoscale information carrier for reversible computing remains elusive.

Magnetic skyrmions [6] are intriguing information carriers for reversible computing due to their small (~50 nm at room temperature, below 20 nm at low temperature) diameter [7–10] and the small current required to induce skyrmion motion [11]. These quasiparticles are topologically stable regions of magnetization comprising a central core oriented anti-parallel to the bulk of a magnetic structure [12–14]. Skyrmion motion involves the propagation of magnetization rather than the transport of physical particles, and can be induced by the spin-Hall effect through the application of an electrical current [15]. These skyrmion quasiparticles move not along the axis of an electrical current, but rather deviate from this axis due to the skyrmion-Hall effect, which is equivalent to the Magnus effect [16,17]. The skyrmion-Hall effect is generally deleterious to device functionality, so a track structure as shown in Fig. 1 is often used to suppress the skyrmion-Hall effect and restrict the motion to a single dimension [18–21]. Magnetic skyrmions propagating along ferromagnetic nanowire [22,23] tracks have been proposed for memory storage and individual logic gates [24–29], but the development of a scalable skyrmion computing system has been impeded by the need to directly cascade skyrmion logic gates without control and amplification circuitry that significantly reduces the computing system efficiency. Furthermore, previous skyrmion logic proposals require the continual creation and annihilation of skyrmions, which is an energetically expensive process that requires an external control system.

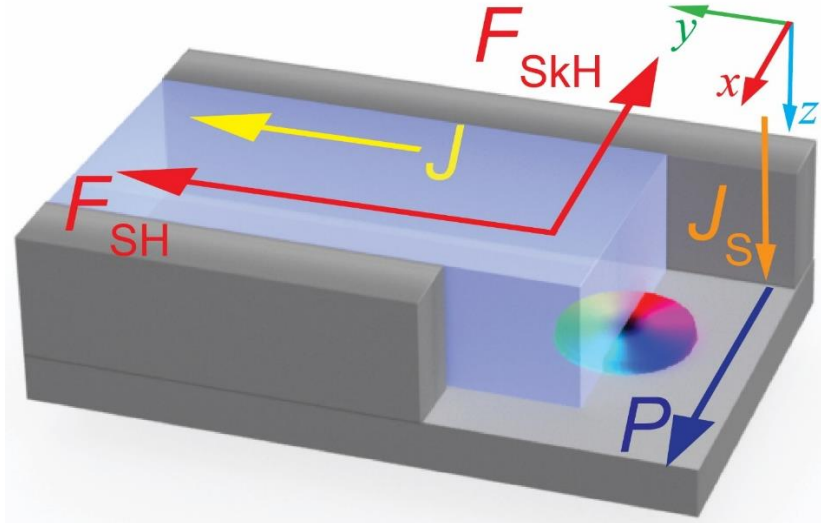

FIG. 1. Skyrmion track structure. Skyrmions propagate along a track comprised of Pt heavy metal (blue) and Co ferromagnet (gray, with polarization $P$), where an interfacial spin-orbit coupling induces a Dzyaloshinskii–Moriya interaction. This spin-orbit coupling also causes the externally-applied electrical current ($J$) flowing through the heavy metal in the +$y$-direction to create a spin current ($J_S$) polarized in the +$z$-direction via the spin-Hall effect. The skyrmion (multicolor circle) lies in the ferromagnetic layer at the interface with the heavy metal, and the surrounding ferromagnet walls prevent the skyrmion from leaving the track. Spin current in the +$z$-direction produces a force $F_{SH}$ on the skyrmions in the +$y$-direction, the direction of the electrical current. The track constriction prevents the –$x$-directed skyrmion-Hall force $F_{SkH}$ from influencing the skyrmion trajectory. Note that the axes are inverted for visual clarity; the Pt heavy metal is below the Co ferromagnet.

We therefore propose a reversible skyrmion logic system in which skyrmions are conserved as they flow through nanowire tracks. Logical operations are performed by thoroughly leveraging the rich physics of magnetic skyrmions: the spin-Hall effect [15], the skyrmion-Hall effect [16,17,30], skyrmion-skyrmion repulsion [31], repulsion between skyrmions and the track boundaries [31], and electrical current-control of notch depinning [11]. Binary information is encoded by the presence ('1') or absence ('0') of magnetic skyrmions, with the skyrmions flowing directly from the output nanowire track of one logic gate to the input track of another logic gate without an external control or amplification circuit. These reversible skyrmion logic gates can provide fan-out and be integrated into a large-scale system, with signal integrity provided by simple electronic clock pulses applied to the entirety of the system. This logic-in-memory computing system is non-volatile due to the topological stability and ferromagnetic nature of skyrmions, providing efficient pipelining that enhances the potential for high speed and low power. Furthermore, the availability of a Fredkin gate inspires the consideration of quantum computing with magnetic skyrmions.

## II. MICROMAGNETIC SIMULATION METHODOLOGY

### A. Micromagnetic Simulation Technique

The Landau-Lifshitz-Gilbert (LLG) equation of motion describes magnetization dynamics in ferromagnetic materials:

$$\frac{\partial \mathbf{M}}{\partial t} = -\gamma(\mathbf{M} \times \mathbf{H}_{\text{eff}}) + \frac{\alpha}{M_s}\left(\mathbf{M} \times \frac{\partial \mathbf{M}}{\partial t}\right) + \boldsymbol{\tau}_{\text{CPP}} \qquad (1)$$

where $\mathbf{M}$ is the magnetization vector, $\gamma$ is the gyromagnetic ratio, $M_s$ is the saturated magnetization and $\alpha$ is the Gilbert damping parameter. $\mathbf{H}_{\text{eff}}$ is the effective field which includes exchange, anisotropy, magnetostatic, Dzyaloshinskii-Moriya and external magnetic fields. $\boldsymbol{\tau}_{\text{CPP}}$ implements the injection of spin-Hall current perpendicularly to the sample and is described by:

$$\boldsymbol{\tau}_{\text{CPP}} = -\beta\epsilon'(\mathbf{M} \times \mathbf{m}_{\text{P}}) - \frac{\beta}{M_s}\big(\mathbf{M} \times (\mathbf{m}_{\text{P}} \times \mathbf{M})\big) \qquad (2)$$

with $\mathbf{m}_{\text{P}}$ as the spin-Hall polarization direction and $\beta = \frac{\theta_{\text{SH}}\hbar J}{2M_s e t_{\text{Co,track}}}$, where $\theta_{\text{SH}}$ is the spin-Hall angle, $e$ is the electronic charge, $J$ is the electrical current density, and $t_{\text{Co,track}}$ is the thickness of the Co track. $\epsilon'$ is the field-like torque, which is here considered to be zero.

Simulations were performed using mumax3, an open-source GPU-accelerated micromagnetic simulation software [32], which integrates the LLG equation of motion with a Finite Difference approach. We discretized the sample into cuboid cells whose dimensions were set to 1 nm × 1 nm × 0.4 nm and we neglected thermal fluctuations by setting the temperature to 0 K. The zero temperature simulation results are predictive of room temperature experimental behavior of the proposed computing system in which a randomly-fluctuating thermal field would impact the skyrmion motion; as described in section V.A, notch synchronizers are used to prevent this random motion from causing logical errors. Furthermore, these phenomena have been suppressed experimentally in multilayer structures [33], and the effect of this thermal field would be small relative to the applied electrical current.

**B. Magnetic Parameter Selection**

The following numerical values – adopted from Purnama et *al.* [18] – model a multilayer of Pt and Co: saturation magnetization $M_s = 5.80 \times 10^5$ A/m, exchange stiffness $A_{ex} = 1,5 \times 10^{-11}$ J/m, Gilbert damping coefficient $\alpha = 0.1$, DMI constant $D_{ind} = 3.0 \times 10^{-3}$ J/m$^2$, magneto-crystalline anisotropy constants $Ku_1 = 6 \times 10^5$ J/m$^3$ and $Ku_2 = 1.5 \times 10^5$ J/m$^3$, and spin polarization in the transverse direction $\boldsymbol{m}_P = (1,0,0)$. The anisotropy direction points upwards. The spin-Hall angle $\theta_{SH}$ is considered to be equal to 1. The thickness of the Pt layer is $t_{Pt} = 0.4$ nm and the thickness of the Co layer varies between $t_{Co,track} = 0.4$ nm (Fig. 1) and $t_{Co,sample} = 0.8$ nm elsewhere. As the skyrmions travel in lithographically-feasible 20 nm-wide nanowire tracks [34], they are confined to slightly smaller dimensions in a manner similar to the confinement shown in [9], where geometric confinement resulted in skyrmions with diameters of roughly 50 nm.

## III. REVERSIBLE SKYRMION LOGIC GATES

### A. Reversible Skyrmion AND/OR Gate

The reversible AND/OR logic function described in Table I is the primary workhorse of the proposed skyrmion computing system. As this function is reversible, the total number of skyrmions $N$ provided to the inputs A and B is always equal to the total number of skyrmions emitted by the AND and OR outputs. This reversible logic function is performed by the structure shown in Fig. 2, with micromagnetic simulations [32] depicting the skyrmion trajectories for each input combination. The spin-Hall effect pushes the skyrmions in the $+y$-direction through the vertical tracks of Fig. 1, while the skyrmion-Hall effect introduces a $-x$-directed force that is mediated by repulsion from the track boundaries. The skyrmions are therefore free to move laterally within the central junction, where the skyrmion-Hall effect causes leftward skyrmion propagation unless repulsed by a second skyrmion.

Table I. Truth table for the AND/OR gate.

| Inputs | | $N$ | Outputs | |
|---|---|---|---|---|
| A | B | | AND | OR |
| 0 | 0 | **0** | 0 | 0 |
| 0 | 1 | **1** | 0 | 1 |
| 1 | 0 | **1** | 0 | 1 |
| 1 | 1 | **2** | 1 | 1 |

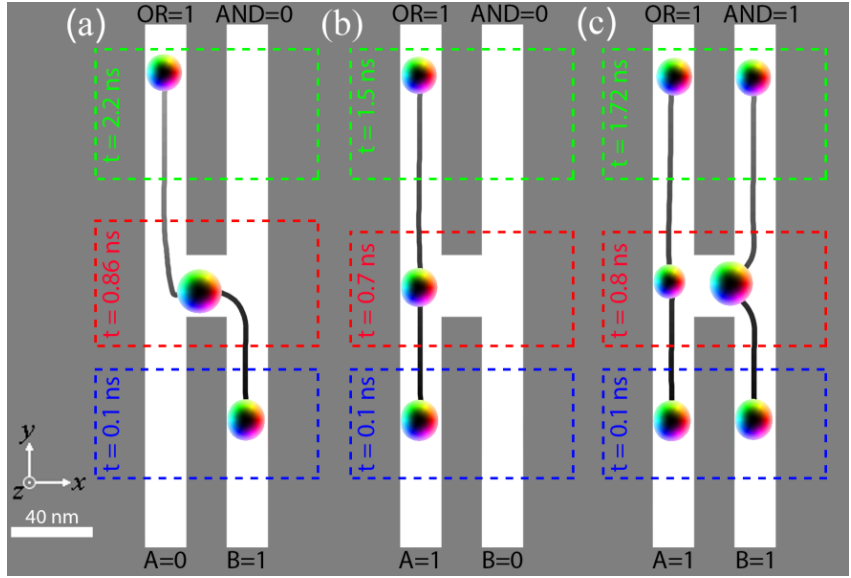

FIG. 2. Micromagnetic simulation results for the AND/OR gate are shown for input combinations (a) A = 0, B = 1; (b) A = 1, B = 0; and (c) A = B = 1. In this and other figures, the path of each skyrmion is shown as a grayscale line that is black at $t = 0$ and gradually lightens as the simulation advances, with skyrmion snapshots provided at the times noted in the figure. The spin current $J_S$ resulting from constant electrical current $J = 5 \times 10^{10}$ A/m² pushes the skyrmions in the +y-direction, with a skyrmion-Hall force directed in the –x-direction. Therefore, when confined laterally by their tracks, the input skyrmions travel directly in the +y-direction until they reach the central junction. In the lateral opening of the constrictive tracks at the central junction, the skyrmion-Hall force induces a –x-directed component to the skyrmion trajectory that is counteracted by skyrmion-skyrmion repulsion.

Whenever a skyrmion enters one of the input ports of the AND/OR gate, this logic gate geometry forces a skyrmion to propagate to the OR output port to represent binary '1'. If two skyrmions are input to this logic gate, one skyrmion is emitted by the OR output port and the other skyrmion is emitted by the AND output port such that both produce binary '1'. Finally, if no skyrmions enter either input port, then no skyrmions are emitted by either output port, representing binary '0' outputs. The combined forces resulting from the spin-Hall effect, the skyrmion-Hall effect, skyrmion-skyrmion repulsion, and repulsion between the skyrmions and the boundaries thus cause this structure to simultaneously calculate the logical functions A ∨ B and A ∧ B while conserving the skyrmions.

**B. Reversible Skyrmion INV/COPY Gate**

The inclusion of an inversion operation enables the generation of all possible Boolean logic functions, which cannot be achieved by the AND and OR operations alone. The proposed INV/COPY gate shown in Table II, and Fig. 3 functions similarly to the AND/OR gate shown above, here with an additional output port and the requirement that a skyrmion always be provided to the control (CTRL) input. This reversible INV/COPY gate simultaneously duplicates and inverts the skyrmion input signal. As shown in Table II, the NOT output is therefore '1' whenever the IN input is '0', and '0' whenever the IN input is '1'. This reversible logic gate also performs the fan-out function, through which skyrmions are conserved such that the IN signal is duplicated to the two COPY outputs. This signal duplication is an essential component of a large-scale computing system, and can be performed repeatedly by cascaded INV/COPY gates to generate numerous copies of a signal.

Table II. Truth table for the INV/COPY gate.

| Inputs | | $N$ | Outputs | | |
|---|---|---|---|---|---|
| CTRL | IN | | COPY1 | NOT | COPY2 |
| 1 | 0 | **1** | 0 | 1 | 0 |
| 1 | 1 | **2** | 1 | 0 | 1 |

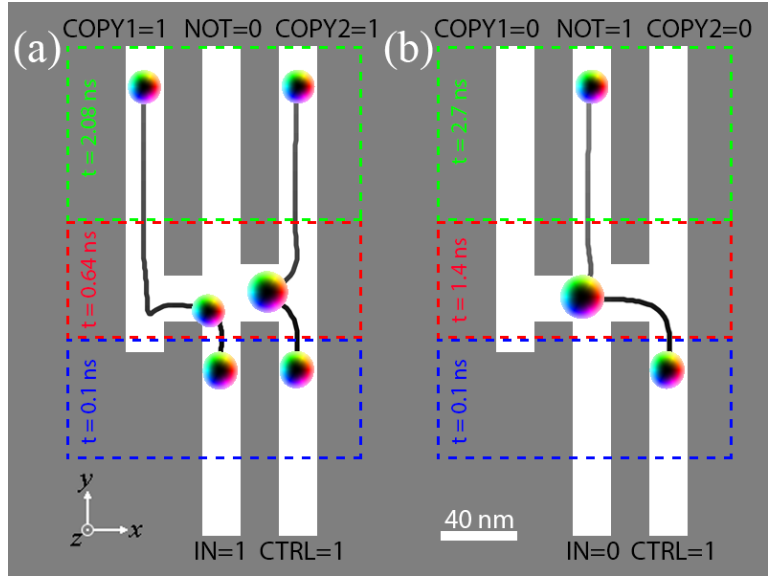

FIG. 3. Micromagnetic simulation results for the INV/COPY gate are shown for input combinations (a) IN = 1, CTRL = 1; and (b) IN = 0; CTRL = 1.

**IV. POTENTIAL FOR SKYRMION QUANTUM COMPUTATION**

A magnetic skyrmion quantum computing system can be envisioned in which quantum non-binary operators operate in concert with the binary operations of a skyrmion Fredkin gate. In particular, the availability of atomic-scale skyrmions [7,35] and the fact that skyrmions coupled to conventional superconductors support Majorana fermions [36] open a pathway to topological quantum computation [37,38]. Building off our proposed skyrmion Fredkin gate and various potential physical solutions for qubit encoding and manipulation, we propose two quantum gate sets for a universal skyrmion quantum computing system: one based on the conventional combination of Toffoli and Hadamard gates, and the other based on Clifford and T gates.

**A. Skyrmion Fredkin Gate**

In addition to reversible Boolean computation, this skyrmion logic system has potential applications in quantum computation. As single-spin states are promising candidates for qubits, the recent experimental demonstration of atomic-scale magnetic skyrmions provides a potential pathway to quantum computing with skyrmion qubits [7,35]. We therefore propose in Fig. 4 and Table III the reversible Fredkin gate implemented in this skyrmion logic paradigm. Here, skyrmions provided to the C input propagate to the C output and determine whether or not the $I_1$ and $I_2$ input signals are swapped as they travel to the $O_1$ and $O_2$ outputs. A controlled-not (CNOT) gate can also be achieved, and is implicitly included in the full adder described in section V.B (Fig. 6(a)).

Table III. Truth table for the Fredkin gate.

| Input Signals | | | $N$ | Output Signals | | |
|---|---|---|---|---|---|---|
| C | $I_1$ | $I_2$ | | C | $O_1$ | $O_2$ |
| 0 | 0 | 0 | **0** | 0 | 0 | 0 |
| 0 | 0 | 1 | **1** | 0 | 0 | 1 |
| 0 | 1 | 0 | **1** | 0 | 1 | 0 |
| 0 | 1 | 1 | **2** | 0 | 1 | 1 |
| 1 | 0 | 0 | **2** | 1 | 0 | 0 |
| 1 | 0 | 1 | **3** | 1 | 1 | 0 |
| 1 | 1 | 0 | **3** | 1 | 0 | 1 |
| 1 | 1 | 1 | **4** | 1 | 1 | 1 |

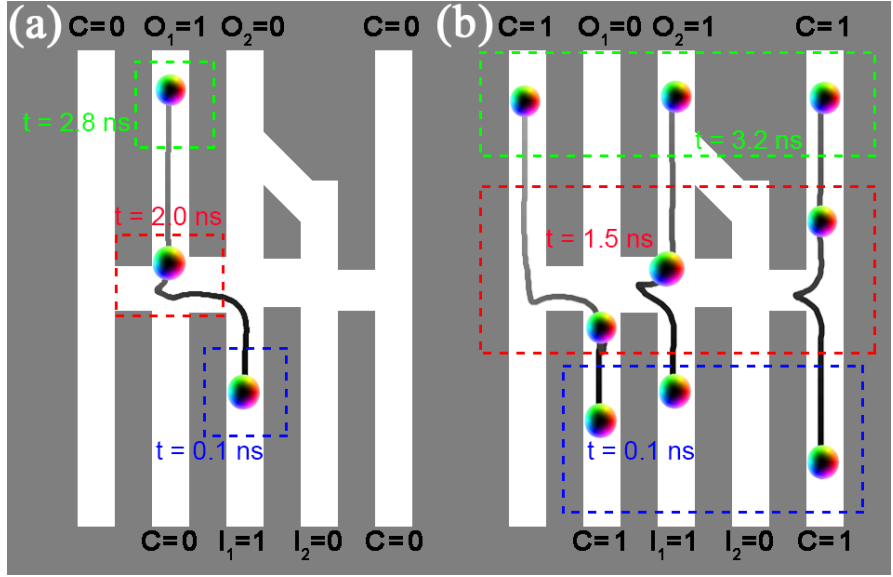

FIG. 4. Micromagnetic simulation results for the Fredkin gate. (a) When C = 0, the input signals travel directly to the output ports without swapping; that is, $O_1 = I_1$ and $O_2 = I_2$. (b) When C = 1, the input skyrmions swap paths such that $O_1 = I_2$ and $O_2 = I_1$. The C signal is duplicated, with both skyrmions at the C input propagating directly to the C output.

## B. Toffoli + Hadamard Universal Quantum Gate Set

The Fredkin gate is an intermediary between classical Boolean logic and full/universal quantum logic, and three Fredkin gates can be cascaded to realize a Toffoli gate. As a Toffoli gate in concert with a Hadamard gate are quantum computational universal, the only missing component required for quantum computing within this skyrmion logic system is the Hadamard gate. The availability of multiple types of skyrmions (antiferromagnetic, etc.) opens a pathway to achieve a skyrmion Hadamard gate analogous to an optical beam splitter with half probability. Here, when a "gate skyrmion" interacts with a "qubit skyrmion," the "gate skyrmion" changes the state of the "qubit skyrmion." The "gate skyrmions" represents the unitary transformation of the quantum Hadamard gate, creating an equal superposition between two possible positions of a particle (the "qubit skyrmion"). "Qubit skyrmions" may be left initialized in '0' or '1' states represented by the absence or presence of a skyrmion, respectively, as described throughout the text. This Hadamard gate thus provides the necessary quantum state superposition.

**C. Clifford + T Universal Quantum Gate Set**

Given their high number of spins and their intrinsic topological properties, skyrmions are intrinsically quantum error corrected logical qubits. It may therefore be feasible to consider the quantum universal Clifford+T gate set [39] in which all skyrmions are qubit carriers. The single qubit T gate can act on individual skyrmions (qubits), while the Clifford gates (efficient to simulate on a classical computer) are implemented by interacting the skyrmions [36]. The most observed and investigated skyrmions are two-dimensional, suggesting that these Clifford gates may be implemented by braiding the movement trajectories of the skyrmions in a manner similar to conventional surface quantum error correcting code [40].

**V. LARGE-SCALE INTEGRATED SKYRMION COMPUTING SYSTEM**

While these reversible logic gates are interesting in their own right, a mechanism for cascading logic gates is necessary for practical computing applications. In the proposed reversible computing system, the output skyrmions emitted by one logic gate are used as input skyrmions for another gate. As the logic gate functionality is based on skyrmion interactions at the central junctions, a synchronization mechanism must also be provided to ensure that skyrmions arriving from different input paths reach the central junction simultaneously, despite thermal effects.

**A. Notch Synchronization with Global Clock**

This synchronization is achieved with the notch structure [11] of Fig. 5(b) by applying a large spin-Hall current pulse that enables the skyrmions to traverse the notch only when this large pulse is applied. This large current pulse causes a decrease in skyrmion diameter, while also increasing the skyrmion velocity as shown in Fig. 5(a). As depicted in Fig. 5(c), a small spin-Hall current is continuously applied to the entire system to propagate the skyrmions through the tracks and logic gate junctions; this current magnitude is below the threshold required for skyrmions to traverse the notches. At regular intervals, a large spin-Hall current pulse is provided to the entire system to drive the skyrmions past the notches; this regular pulse represents the global system clock that synchronizes the computing system. These notch synchronizers can be placed following the output of a logic gate, with the IN port of the notch synchronizer connected to an output port of a logic gate; the OUT port of the notch synchronizer is connected to an input port of a cascaded logic gate. Notches are inserted between every logic gate input and output where synchronization is required, with each notch synchronizer handling zero or one skyrmion during each clock cycle. Alternatively, this synchronization can be similarly achieved through clocked electrical control of the magnetic anisotropy [41].

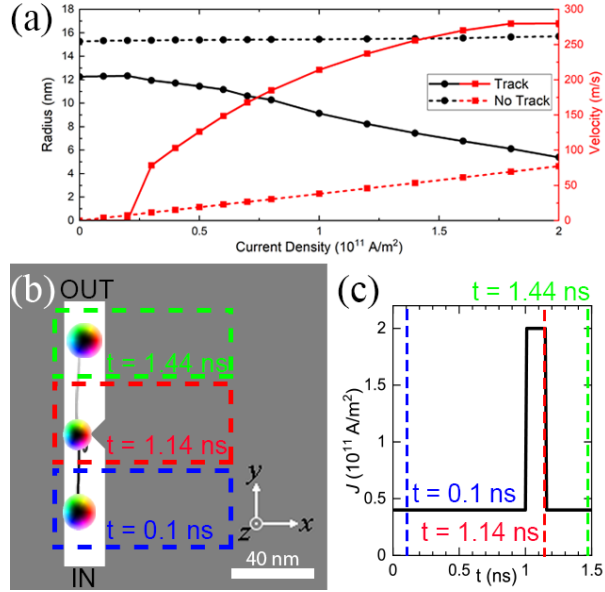

FIG. 5. Signal synchronization. (a) Skyrmion radius as a function of applied electrical current density. (b) A 7 nm-wide notch is formed in the 20 nm-wide nanowire track to create a constriction that permits skyrmion passage only when a large current is applied. (c) The electrical current applied to the entirety of the computing system maintains a constant low magnitude of $J = 5 \times 10^{10}$ A/m² that is periodically amplified to $J = 2 \times 10^{11}$ A/m² for 150 ps in order to enable skyrmions to traverse notches throughout the system. The skyrmion traverses the notch when this large clock pulse is applied at $t = 1$ ns.

### B. Directly-Cascaded Skyrmion Logic Gates

Integrating the basis logic gates with the cascading and synchronization mechanisms enables the scaling of this reversible computing paradigm to large systems that efficiently perform complex functions. An example is provided in Fig. 6, where the input A, B, and carry-in skyrmion signals interact as they propagate through the circuit to produce the sum and carry-out skyrmion signals, thus executing the one-bit full addition function with two half adders. A 150 ps-wide clock pulse is provided every 5 ns to synchronize the skyrmions, ensuring proper conservative logic interactions within each component logic gate. The sum output is produced within three clock cycles, while the carry-out output is produced within two clock cycles; the carry-in to carry-out delay is only one clock cycle. These clocked skyrmion signals provide a natural means for pipelining, enabling the execution of *n*-bit addition within *n*+2 clock cycles; this pipelining procedure can be observed in Fig. 7. Furthermore, though the 200 MHz clock frequency and the electrical current magnitudes used in simulation provide inferior efficiency as compared to conventional computing systems, the non-volatility and pipelining inspire a vision for highly-efficient computing with alternative materials [42] and improvements in the Rashba coefficient and spin-Hall angle [43].

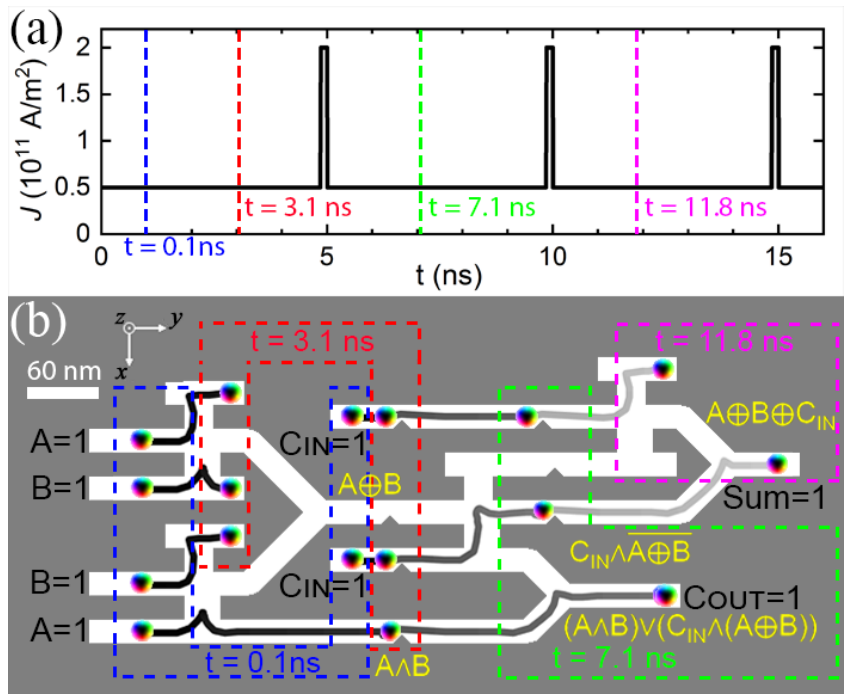

FIG. 6. Cascaded one-bit full adder. (a) Cascaded logic gates are synchronized by large 150 ps-wide electrical pulses applied to the entire structure with a clock period of 5 ns. (b) A one-bit full adder computes the binary sum and carry-out of two one-bit binary numbers A and B and a carry-in bit. The notches ensure that the skyrmions are synchronized with one another as they enter each logic gate, thereby providing proper skyrmion-skyrmion repulsion and logical functionality. After two (three) clock cycles, the skyrmion signals reach the $C_{OUT}$ (SUM) outputs.

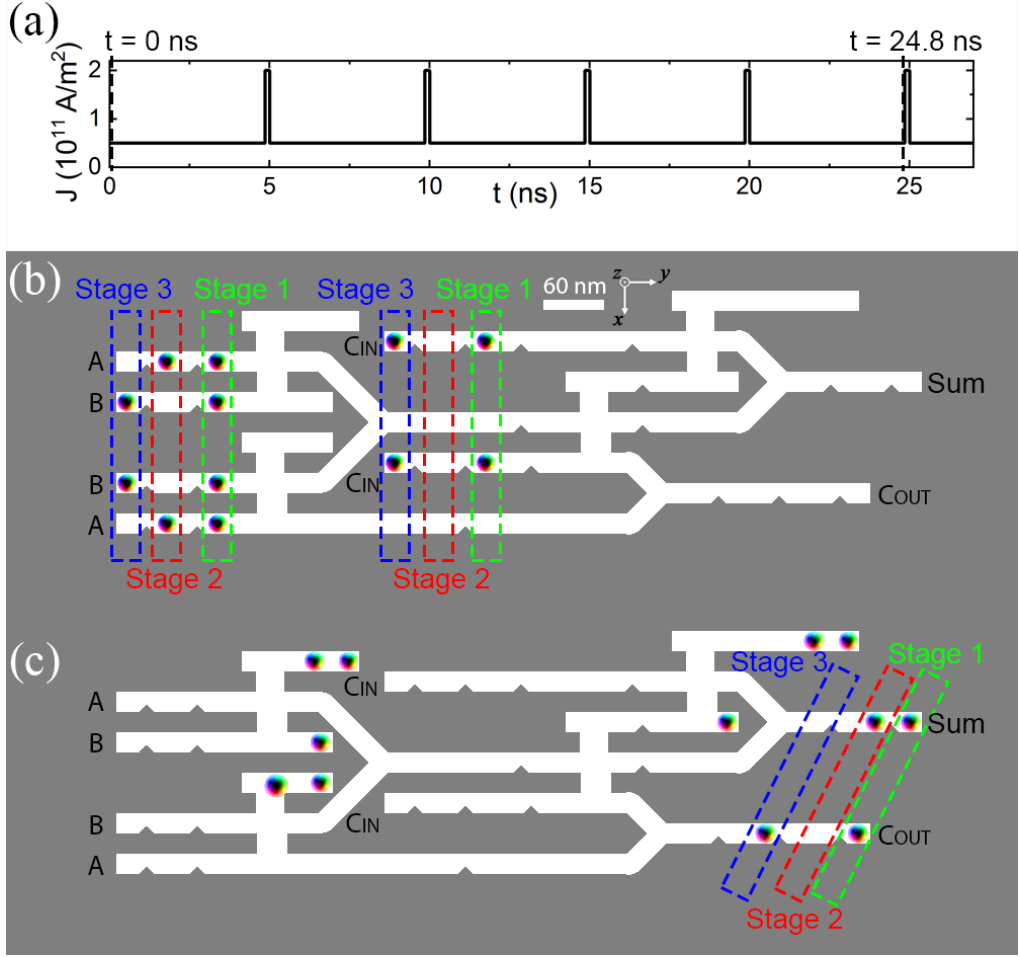

FIG. 7. Pipelined full adder. The clocked full adder structure can be used for pipelined operations such that each logic gate junction is used within each clock cycle. (a) The clocking scheme used in the full adder micromagnetic simulation is extended for five cycles to perform three separate full adder operations simultaneously. (b) The initial state (t = 0 ns) of a three-stage pipeline is shown here for the following inputs: In stage one, A = B = $C_{IN}$ = 1. In stage two, A = 1 and B = $C_{IN}$ = 0. In stage three, A = 0 and B = $C_{IN}$ = 1. (c) The final state (t = 24.8 ns) is seen after five clock cycles, with the following results: For stage one, Sum = $C_{OUT}$ = 1. For stage two, Sum = 1 and $C_{OUT}$ = 0. For stage three, Sum = 0 and $C_{OUT}$ = 1.

## VI. EXPERIMENTAL APPROACHES

The skyrmion logic system proposed here is experimentally feasible with industrially relevant magnetic multilayers of the required material systems at lithographically-accessible length scales. Furthermore, room temperature field-free operation is achievable, as nanoscale chiral skyrmions were recently stabilized at room temperature with zero magnetic field [33,44,45]. Similar to the material parameters used in the above micromagnetic simulations, a Co-based multilayer (Ir/Pt/Co) stack, grown by sputtering, has been proposed that should enable room temperature sub-100 nm skyrmions stabilized by a large additive Dzyaloshinskii–Moriya interaction (DMI) [33]. By further tuning material parameters such as the perpendicular anisotropy, the thickness of the magnetic layers, or the DMI, even smaller skyrmion dimensions can be achieved [11]. Full-field X-ray magnetic imaging techniques can be used to demonstrate the device functionality and logic operations, including full-field magnetic transmission soft X-ray microscopy (MTXM) or photoemission electron microscopy (PEEM), which is a full-field surface sensitive X-ray imaging technique. X-ray imaging is magnetization-sensitive and non-disruptive to the sample. For the proposed system in particular, imaging of the magnetic states of the devices must utilize the X-ray magnetic circular dichroism (XMCD) effect and tune the incoming photon energy to the Co L3 edge (~779 eV).

**A. Skyrmion Generation**

In a fully-reversible skyrmion logic system, skyrmions must only be generated once, at the beginning of the system operation. After these initial skyrmions are generated, they are continually propagated through the logic gates such that the output skyrmions of each conservative skyrmion logic gate are used as the input skyrmions of other conservative skyrmion logic gates (see section VI.C). These skyrmions generated at system initialization are therefore sufficient for long-term use of this system, and their non-volatility enables the skyrmions to maintain their states even when the power supply is removed.

As skyrmion generation is significantly more challenging than the generation of many other types of information carriers, this ability to conserve skyrmions marks a major advance in comparison to previous proposals for skyrmion logic [24–29]. As this reversible computing system only requires a single skyrmion initialization process, the precise technique used has limited impact on the overall system efficiency. Several techniques are available, with the application of nanosecond electrical current pulses [8] appearing particularly promising.

Alternatively, ease-of-demonstration as well as optimization of the conventional metrics of speed, power, and delay may inspire a compromise regarding the conservation of skyrmions. Another approach to skyrmion generation is to continually generate skyrmions at specific points within the system with a dedicated skyrmion generation structure. This could be achieved with homogeneous currents [8], or with nanosecond electrical current pulses as achieved in [46] for a device-compatible stripline geometry. This skyrmion generator would generate a skyrmion in each clock cycle, which then can be provided as an input to a particular logic gate where an input skyrmion is always required (*e.g.*, for the CTRL input of the INV/COPY gate). These skyrmions would then travel through the circuit, and eventually be annihilated after they have performed the desired operations. While this alternative system would no longer be fully reversible, the simplicity of this approach may be advantageous for proof-of-concept experiments as well as eventual high-performance computing systems.

**B. Skyrmion Detection**

To read the binary outputs of this conservative logic system, it is necessary to detect the presence (binary '1') or absence (binary '0') at various points throughout the circuit. While standard magnetic force microscopy imaging can be used to detect skyrmions in a laboratory setting, computing applications require transformation of the skyrmion information into electrical signals. Determination of the presence or absence of a skyrmion at a particular location can be achieved by placing a tunneling barrier and hard ferromagnet above the Co ferromagnet to form a magnetic tunnel junction (MTJ), as has been described in previous domain wall and skyrmion logic and memory proposals [25,27,29]. The magnetoresistance of this MTJ indicates the presence or absence of a skyrmion within the free layer, as the skyrmion modifies the local magnetization within the free layer and therefore the current through the MTJ tunneling barrier.

Skyrmion Conservation

While each of the skyrmion logic gates conserves skyrmions by propagating each input skyrmion to an output port, the operation of a complete conservative logic system requires the conservation of skyrmions throughout the system. It is therefore critical that every skyrmion transmitted to the output port of a skyrmion logic gate is then used as an input to another skyrmion logic gate.

As can be observed in relation to the full adder of Fig. 6, skyrmions propagate to several superfluous output ports that contain logical by-products of the computation of the sum and carry-out signals. In particular, the full adder also produces the signals A, B, A∧B, $C_{IN}$, and $C_{IN} \wedge (A \oplus B)$; these are by-products of the full adder computation that are not the primary objectives of the full adder circuit. These signals are therefore available for use in other logic gates; it may be noted that one signal is available at the output for each of the three input signals.

To enable a reversible system, these skyrmion signals must be able to propagate to other logic gates. As the lateral flow of information through this two-dimensional structure is unidirectional (left-to-right), it is necessary to provide a technique by which skyrmions are provided to the left side of the circuit. Many potential techniques may be available, such as the use of an additional circuit layer, which would also enable the interaction-free cross-over of skyrmion tracks. Furthermore, the direction of the spin current can be modified through static or dynamic modulation of the electrical current direction or the ferromagnetic Co magnetization. Finally, it should be noted that it may be worthwhile to shed the requirement of complete skyrmion conservation in order to maximize the primary metrics of a computing system (energy consumption, processing speed, and area footprint).

## VII. CONCLUSIONS

The proposed skyrmion structure enables the realization of a nanoscale reversible computing system with nearly-ideal characteristics. Small spin-Hall currents are supplied and the energy required for motion approaches the reversible computing ideal of frictionless information propagation. Rather than avoiding the skyrmion-Hall effect, the proposed system leverages the rich physics of magnetic skyrmions to provide cascaded logic operations that are pipelined and synchronized to maintain signal integrity when scaled to large systems. In addition to Boolean logic, the availability of a Fredkin gate inspires a vision for quantum computing with magnetic skyrmions. Furthermore, the stability of the ferromagnetic materials provides non-volatility that can be exploited in non-von Neumann architectures that overcome the limitations of conventional computing systems. By performing conservative logic with magnetic skyrmions, the proposed system enables a reversible computing paradigm that approaches the ultimate thermodynamic limits of information processing efficiency.

AUTHOR INFORMATION

**Corresponding Author**

*E-mail: joseph.friedman@utdallas.edu.

ACKNOWLEDGMENTS

The authors thank H.-J. Drouhin for arranging the research exchange and M. R. Stan for fruitful discussions.